\documentstyle[twoside,psfig]{article}
\input{ijmpb-macros}

\begin{document}

\runninghead{Nonadiabatic theory of the superconducting state} 
{Nonadiabatic theory of the superconducting state}

\normalsize\textlineskip
\thispagestyle{empty}
\setcounter{page}{1}

\copyrightheading{}                     

\vspace*{0.88truein}

\fpage{1}
\centerline{\bf NONADIABATIC THEORY OF THE SUPERCONDUCTING STATE}
\vspace*{0.237truein}

\centerline{\footnotesize MICHELA BOTTI}
\vspace*{0.015truein}
\centerline{\footnotesize\it Dipartimento di Fisica, 
Universit\`a ``La Sapienza'', P.le Aldo Moro 2}
\baselineskip=10pt
\centerline{\footnotesize\it Roma, 00185, Italy} 
\baselineskip=10pt
\centerline{\footnotesize\it and INFM, Unit\`a Roma1}

\vspace*{10pt}
\centerline{\footnotesize EMMANUELE CAPPELLUTI}
\vspace*{0.015truein}
\centerline{\footnotesize\it Dipartimento di Fisica, 
Universit\`a ``La Sapienza'', P.le Aldo Moro 2}
\baselineskip=10pt
\centerline{\footnotesize\it Roma, 00185, Italy} 
\baselineskip=10pt
\centerline{\footnotesize\it and INFM, Unit\`a Roma1}

\vspace*{10pt}
\centerline{\footnotesize CLAUDIO GRIMALDI}
\vspace*{0.015truein}
\centerline{\footnotesize\it \'Ecole Polytechnique F\'ed\'erale, 
D\'epartment de microtechnique IPM}
\baselineskip=10pt
\centerline{\footnotesize\it Lausanne, CH-1015, Switzerland}

\vspace*{10pt}
\centerline{\normalsize and}

\vspace*{10pt}
\centerline{\footnotesize LUCIANO PIETRONERO}
\vspace*{0.015truein}
\centerline{\footnotesize\it Dipartimento di Fisica, 
Universit\`a ``La Sapienza'', P.le Aldo Moro 2}
\baselineskip=10pt
\centerline{\footnotesize\it Roma, 00185, Italy} 
\baselineskip=10pt
\centerline{\footnotesize\it and INFM, Unit\`a Roma1}
\vspace*{0.225truein}

\abstracts{Fermi energies in fullerene compounds and cuprates
are extremely small
as consequence of the small number of charge carriers and are comparable
to the phonon frequency scale. In this situation the conventional
Migdal-Eliashberg theory does not hold anymore and nonadiabatic effects need
to be taken into account. In previous studies, a generalization
of Eliashberg theory in the nonadiabatic regime has been proposed
to calculate normal state properties and the
onset temperature $T_c$ of the superconductive phase.
Here we extend the nonadiabatic theory below $T_c$ where the opening
of the superconducting order parameter affects the nonadiabatic correction.
The superconducting gap $\Delta$ is calculated in a self-consistent way.
We find that large values of the ratio $2 \Delta/T_c$ are obtained
in the nonadiabatic theory by smaller electron-phonon coupling $\lambda$
than in Migdal-Eliashberg theory. This agrees with the picture
that strong-coupling phenomenology can be achieved in nonadiabatic theory
by ``reasonable'' values of $\lambda$. We apply our analysis
to the case of the fullerene compounds.}{}{}

\textlineskip                   
\vspace*{12pt}                  

\textheight=7.8truein
\setcounter{footnote}{0}
\renewcommand{\thefootnote}{\alph{footnote}}

\section{Introduction}
A common peculiar characteristic of many unconventional
high-$T_{c}$ superconductors (cuprates, 
A$_{3}$C$_{60}$ compounds,...) is the narrowness of the electronic
bands crossing the Fermi level, leading to Fermi energies $E_F$
of the same order of magnitude of the phonon energies $\omega_{\rm ph}$
($\omega_{\rm ph}/E_F \div$ 0.2 - 0.3).
In such a situation, the adiabatic assumption 
($\omega_{\rm ph}/E_F \ll$ 1) on which
Migdal's theorem\cite{migdal} relies cannot be used anymore
to justify the omission of vertex corrections in a diagrammatic theory.
Motivated by this observation, a renewed interest has recently
arose about the possible effects due to inclusion of 
the vertex corrections in the electron-phonon (or any kind
of bosonic mediator) interaction.
This task is even more relevant for the purely electronic interaction models
(Hubbard, or $t-J$), where no justification 
at all
to neglect such corrections exists.

In this perspective, in the past years, we have performed an intensive study
of the vertex corrections in the normal state, and, afterwards,
we have generalized the conventional Migdal-Eliashberg theory in order to
include the first nonadiabatic corrections due to the breakdown of Migdal's 
theorem \cite{psg,gps,gpsprl}. We have investigated the effect of the
Migdal corrections on different quantities related to both the
superconducting and the normal state properties, as for instance the 
superconducting critical temperature and its isotope effect $\alpha_{T_c}$, or 
the isotope effect on the effective electronic mass $\alpha_{m^*}$ \cite{isot}.
Significative nonadiabatic effects have been predicted
on the electronic heat capacity, the spin susceptibility,
the reduction rate of $T_c$ by impurity scattering, and they can
be used as experimental tests of the nonadiabatic theory.

So far the nonadiabatic theory has focused on the analysis
of normal state properties and of the superconducting transition
($T_c$, $\alpha_{T_c}$, \ldots) which can be calculated
as instability of the normal state. However, a generalization
of the nonadiabatic theory under $T_c$, in the superconducting state,
is required in order to investigate nonadiabatic effects
on superconductive quantities as the gap or the penetration depth.
In particular, the finite and negative isotope effect on the electronic mass
$m^*$ has been experimentally inferred by measurements of penetration
depth in the superconducting state \cite{zhao}. It would be therefore
of the highest interest to build up a nonadiabatic theory of the
superconducting state and compare the isotope effect on the
penetration depth at zero temperature with the predicted isotope
effect on $m^*$ evaluated in the normal state.
Self-consistent calculation of the superconducting gap $\Delta$ in
nonadiabatic regime is also of primary relevance.
A recent study of the experimental scenario of Rb$_3$C$_{60}$,
based on the crossed analysis of $T_c$, $\alpha_{T_c}$ and $\Delta$,
points out that the conventional Migdal-Eliashberg theory fails to describe 
superconductivity in the fullerides. On the other hand, preliminar 
investigations, restricted to $T_c$ and $\alpha_{T_c}$, suggest that the 
nonadiabatic theory could account much more properly for the superconducting 
properties \cite{rbc}. In order to complete such analysis and compare the 
results with the experimental ones, a self-consistent calculation
of the gap in nonadiabatic theory is therefore needed.

In this contribution we outline the main ingredients of a nonadiabatic
theory of the superconducting state. As first step we investigate
the modification induced by the superconducting gap opening
on the vertex corrections arising from the breakdown of Migdal's theorem.
Afterwards a generalization of the nonadiabatic equations in the
superconducting state will be constructed.

\section{Vertex Corrections in the Superconducting State}

Migdal's theorem is a basic assumption in the conventional
theory of superconductivity and can be considered the equivalent of 
Born-Oppenheimer approximation in quantum field theory.
The electronic self-energy $\Sigma$ due to the phonon interaction can be 
expressed in exact way as:
\begin{equation}
\Sigma(k)= - \int d^4q \:G(k) D(q) \Gamma(k,q),
\end{equation}
where $k$ and $q$ are momentum-frequency quadrivectors
and $G$, $D$, $\Gamma$ are respectively the electron and phonon propagators
and the vertex function. Migdal's theorem states the vertex corrections
to the bare vertex are of the order of the adiabatic parameter 
$\omega_{\rm ph}/E_F$:
\begin{equation}
\Gamma = 1+P =
1 + O\left(\frac{\omega_{\rm ph}}{E_F}\right),
\end{equation}
where $\omega_{\rm ph}$ is the characteristic phonon frequency scale
and $E_F$ the Fermi energy. In high-$T_c$ materials, where
$\omega_{\rm ph}/E_F$ is not negligible, the vertex correction $P$
can not be anymore neglected and needs to be explicitely taken into account.

In the past years we have generalized the Eliashberg equations at $T_c$
to include nonadiabatic vertex corrections due to the violation
of Migdal's theorem. A crucial role is played by the
momentum-frequency (${\bf q}-\omega$)
structure of the vertex correction $P({\bf q},\omega)$, schematized by the two
representative limits, the static (${\bf q}\rightarrow 0,\omega=0$)
and dynamic (${\bf q}=0,\omega\rightarrow 0$) one.
Vertex corrections are mainly positive in the dynamic 
($v_F |{\bf q}|/\omega < 1$) and negative in the static 
($v_F |{\bf q}|/\omega > 1$) regime. It is thus clear
that the resulting effect of the vertex corrections will depend on
the global balance between positive and negative parts and can be affected
by the physics of the system. For instance, strong correlation
has been shown to select mainly positive parts\cite{grilli} leading to
an enhanced effective electron-phonon coupling. On the contrary,
non magnetic impurity scattering enlarges the negative part \cite{scatt} 
yielding a reduction of $T_c$ even in $s$-wave superconductivity. A similar 
situation is recovered when the superconducting gap opens under $T_c$. 
In Fig.~1
\begin{figure}
\centerline{\psfig{figure=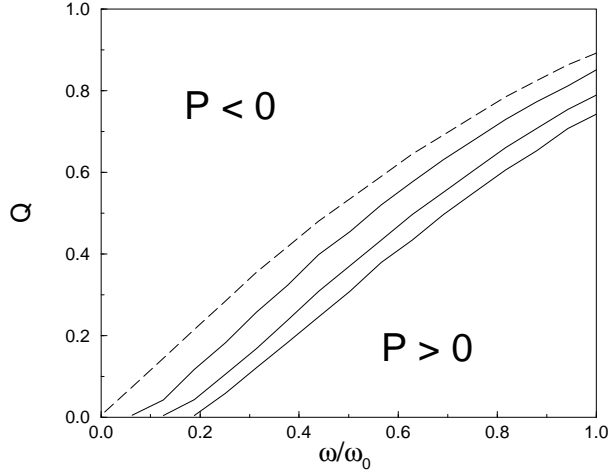,width=8cm}}
\caption{
Plot in the $\omega,Q$ space of the curve on which vertex function
takes zero value; the dashed line corresponds to the normal state, 
solid ones to the superconducting state for different values of
the gap: (from top to the bottom)
$\Delta/\omega_0=0.096$, $\Delta/\omega_0=0.168$, $\Delta/\omega_0=0.217$.}
\end{figure}
the sign of the vertex function in the ($\omega,Q$) space
of the exchanged frequency and momentum ($Q=2|{\bf q}|/k_F$ with $k_F$ the
Fermi vector) is plotted. Solid and dashed lines represent respectively the
$P(Q,\omega)=0$ curve in the superconducting and normal state.
In the latter one the $P(Q,\omega)=0$ curve crosses the point
($Q=0,\omega=0$) where the vertex correction is non-analytical,
in agreement with the static and dynamic limits. As shown in the figure,
the opening of the superconducting gap removes this non-analitical point and
the dynamic and static limit are found to be identical and
both negatives, for $T<T_c$. 

\section{Superconducting Gap and Strong Coupling Phenomenology}

To calculate the superconducting gap in a selfconsistent way, including the
nonadiabatic corrections, we need a set of generalized Eliashberg
equations. The nonadiabatic corrections are constructed, in the
superconducting as in the normal state, whithin a perturbative
scheme based on $\lambda\omega_{\rm ph}/E_F$ parameter ($\lambda$ being the
electron-phonon coupling constant). In order to obtain explicit expressions of
the vertex and cross scattering corrections, we make the following assumptions:
\begin{itemize}
\item a flat density of states in the half filling case

\item an Einstein phonon spectrum 
with frequency $\omega_0$; 

\item the bare electron-phonon vertex is
given by:
\[
|g_{\bf{k},\bf{k}+\bf{q}}|^2=g^2\left(\frac{2\,k_F}{q_c}\right)^2
\theta(q_c-|\bf{q}|)
\]
where $q_c$ is a cut off on the exchanged momenta
due to the effect of electronic
correlations \cite{grilli}.
\end{itemize}

The equations for the wave function renormalization factor $Z_n$ and the gap
function $\Delta_n$, containing the first order corrections beyond
Migdal's limit, for $T<T_c$, can be written as follows:
\begin{eqnarray}
Z_n&=&1+\frac{2 T}{\omega_n}\sum_m
\lambda_Z(Q_c;i\alpha_m,i\alpha_n)\frac{Z_m\omega_m}{\alpha_m}
\arctan\left(\frac{E_F}{\alpha_m}\right)\label{eqel0}
\\
Z_n\Delta_n&=&2\,T\sum_m\lambda_{\Delta}(Q_c;i\alpha_m,i\alpha_n)
\frac{Z_m\Delta_m}{\alpha_m}\arctan\left(\frac{E_F}{\alpha_m}\right)
\label{eqel}
\end{eqnarray}
where $\alpha_n=Z_n\sqrt{\omega_n^2+\Delta_n^2}$ and $\omega_n$ are fermionic
Matsubara frequencies. The terms $\lambda_Z$
and $\lambda_{\Delta}$ represent the
electron-phonon kernel functions calculated at first order
in $\omega_0/E_F$ for
the diagonal and off-diagonal interactions.
It is important to stress that $\lambda_Z$ and $\lambda_{Delta}$ contain different 
nonadiabatic corrections and have therefore different
functional forms. In the 
$T\rightarrow T_c$ limit, equations (\ref{eqel0})-(\ref{eqel})
reduce to the linearized ones, derived in Ref. 3.
In Fig.~2 we show the 
diagrammatic expression of the generalized
\begin{figure}
\centerline{\psfig{figure=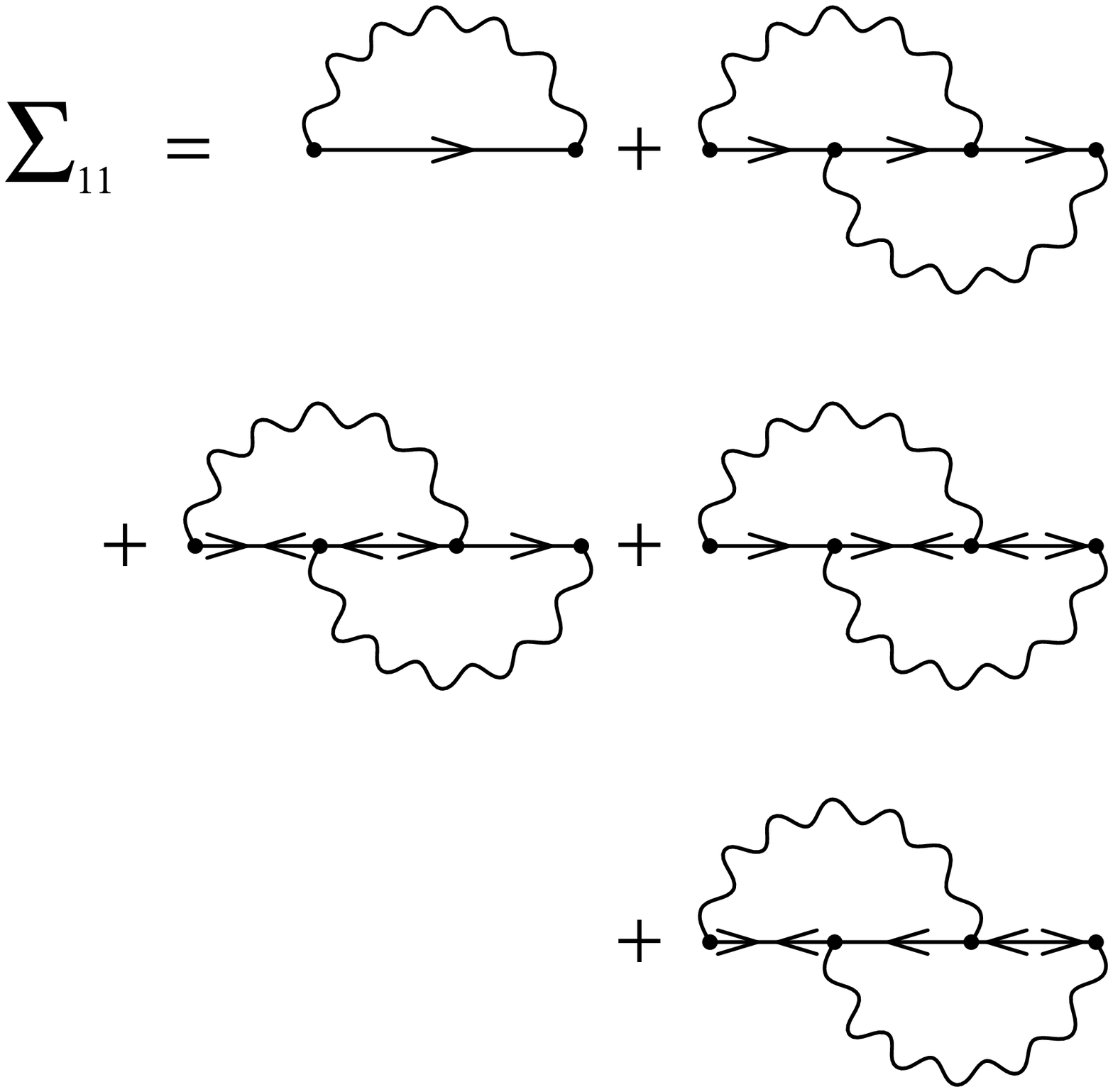,width=6cm,height=6cm}
\psfig{figure=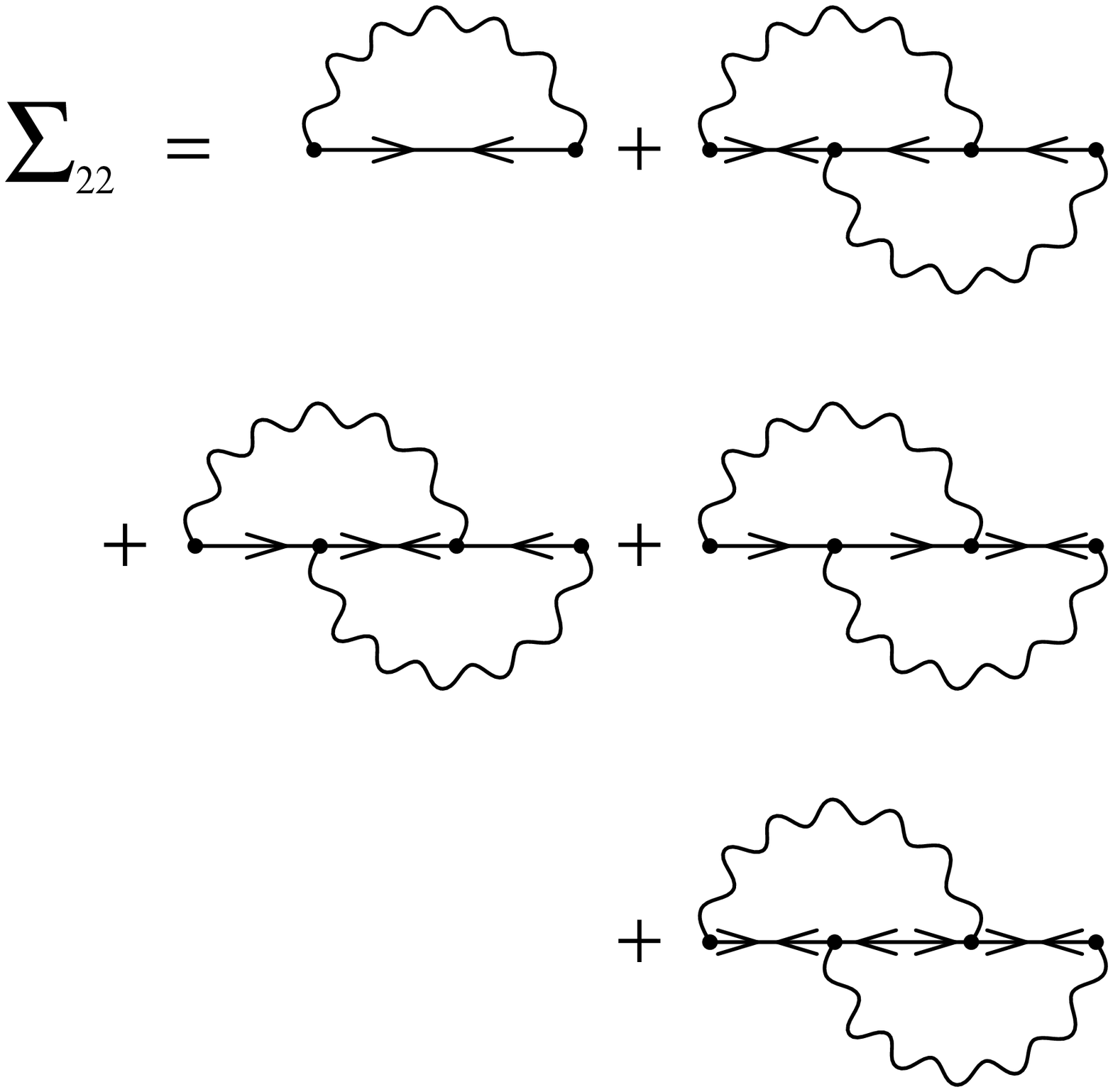,width=6cm,height=6cm}}
\caption{Diagrammatic expression of generalized Eliashberg equations 
in the superconducting state.}
\end{figure}
Eliashberg equations for normal and anomalous self-energy, including the first
nonadiabatic correction, for $T<T_c$. By using this closed set of selfconsistent
equations, one can calculate the wave function renormalization factor
$Z_n$ and the superconducting gap $\Delta_n$, in Matsubara frequency space. 

\begin{figure}
\centerline{\psfig{figure=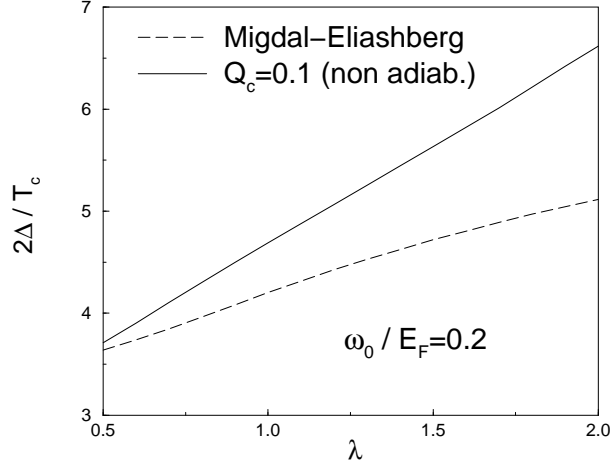,width=8cm,clip=!}}
\caption{Plot of the ratio $2\Delta/T_c$ as function of
$\lambda$; solid line corresponds to the nonadiabatic case
and dashed line to the Migdal-Eliashberg result.}
\label{tcl}
\end{figure}

By solving numerically the generalized Eliashberg equations, we have
investigated the effects of the nonadiabatic corrections on the gap function
and on the $2\Delta/T_c$ ratio. We find that, for a fixed value of $T_c$,
nonadiabatic corrections suppress the gap function because of the enhancement
of the negative vertex region due to the opening of the superconducting gap. 
In this case, however, adiabatic and nonadiabatic solutions correspond to two 
different
values of $\lambda$, of which the larger one is required in Migdal-Eliashberg
theory. For a given $\lambda$, instead, nonadiabatic corrections increase
the value of the gap and the $2\Delta/T_c$ ratio as well 
(see Fig. \ref{tcl}).
To clarify this 
result we note that for a given
value of $\lambda$ nonadiabatic corrections produce a large enhancement of
$T_c/\omega_0$ and of $\Delta/\omega_0$, namely give arise to
a ``strong coupling'' phenomenology defined by a non negligible ratio
between the superconducting energy scale and the phononin
one. For not too large values of $\omega_0/E_F$, this effect is more 
relevant than the gap suppression due to the enhancement of negative contributions
in the vertex corrections. 

An interesting result of the nonadiabatic theory in superconducting state 
is that the 
electron-phonon coupling constant $\lambda$ required to obtain
a given $2\Delta/T_c$ ratio, whithin nonadiabatic framework, is found to be
significantly smaller than the one needed in Migdal-Eliashberg theory 
\cite{rbc}.  

Migdal-Eliashberg theory is often used to describe
the superconducting properties of the 
alkali-doped C$_{60}$ compounds. In order to account for the experimental
critical temperature $T_c=30$ K a very large
microscopical electron-phonon coupling constant ($\lambda > 3$) is needed.
On the other hand, the comparison of phononic and electronic energies
gives compelling evidence of the nonadiabaticity in these
materials. Recent studies show that 
the experimental scenario in Rb$_3$C$_{60}$ ($T_c=30$ K,
$\alpha_{T_c}=0.21$) can be naturally reproduced in the nonadiabatic
theory of superconductivity with reasonable values of $\lambda$
and $\omega_0$\cite{rbc}.
This analysis should be completed by taking into account
also the size of the superconducting gap as estimated by experimental
measurements. To this aim the present results can be used to
determine the values of the superconducting gap $\Delta$ and of
the ratio $2\Delta/T_c$. We have shown that the measured
$2\Delta/T_c=4.2$ corresponds to a
value of the microscopical parameter $\lambda$
which is significantly smaller in nonadiabatic theory than the one
derived by using conventional Migdal-Eliashberg framework,
in agreement with the above discussion.


\end{document}